\documentclass[fleqn,twoside]{article}
\usepackage{espcrc2}

\readRCS
$Id: espcrc2.tex,v 1.2 2004/02/24 11:22:11 spepping Exp $
\usepackage{graphicx}
\usepackage[figuresright]{rotating}

\newcommand{\AmS}{{\protect\the\textfont2
  A\kern-.1667em\lower.5ex\hbox{M}\kern-.125emS}}

\title{Status of Pentaquark Search at Jlab}

\author{Valery Kubarovsky and Paul Stoler
\address[RPI]{Rensselaer Polytechnic Institute,
        Troy, NY 12180}
\\
        For the Jefferson Lab CLAS Collaboration}

\begin{document}

\begin{abstract}
We review the current experimental 
situation of pentaquark searches, 
and second generation experiments, with emphasis 
on the Jefferson Lab program. 
\vspace{1pc}
\end{abstract}

\maketitle

\section{INTRODUCTION}
\begin{figure}[htb]
\vspace{9pt}
\centerline{\includegraphics[width=7cm]{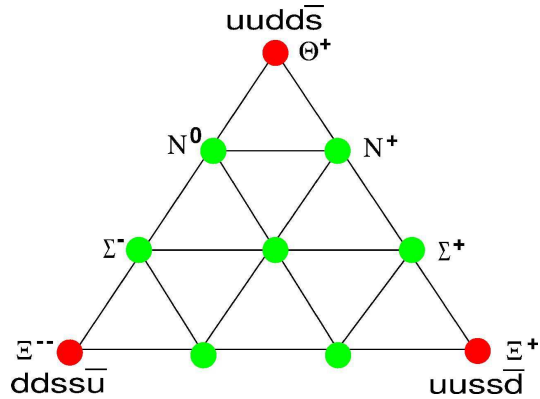}}
\caption{Anti-decuplet of baryons. The corners of this diagram are manifestly exotic.}
\label{fig:antidecuplet}
\end{figure}

The possibility of  the existence of pentaquarks has been theoretically
discussed for many years. However,  the subject has been thrust to the 
forefront  during the past two years by the report of the 
possible observation of a such a state at Spring-8 \cite{LEPS_old}
and ITEP \cite{DIANA}.  This was immediately 
followed up by several positive signals from groups analyzing previously
obtained data \cite{CLAS1,CLAS2,CLAS3,SAPHIR}, 
so that at this writing there are more than a dozen 
observations of a state having mass $M=1522-1555$~MeV, strangeness $S=+1$, (the 
negative of that for normal strange baryons), and very narrow width 
$\Gamma < 10$~MeV 
\cite{LEPS_old,DIANA,CLAS1,CLAS2,CLAS3,SAPHIR,HERMES,ITEP,SVD,COSY,ZEUS}. 
This new state, dubbed $\Theta^+$ \cite{DPG2004},  was 
identified  as a candidate for the $uudd\bar s$ lowest lying member of a
predicted  pentaquark baryon anti-decuplet
\cite{Diakonov97}, illustrated in Fig.~\ref{fig:antidecuplet}. 

In addition to the exotic $\Theta^+$, the anti-decuplet predicts the existence 
of two 
other exotic states which have S=-2 and charge Q=-2 and Q=+1, 
denoted  $\Xi^{--}_5$ and $\Xi^+_5$ respectively. The subscript ``5'' indicates the
five-quark (pentaquark) nature of the states and is used to distinguish them from 
ordinary cascade states. 
These exotic cascade states have isospin 3/2. Two additional partners, denoted
$\Xi^{-}_5$ and $\Xi^0_5$, are also 5-quark states but are not explicitly exotic.
The CERN NA49 collaboration has recently reported evidence for
$\Xi^{--}_5$ and $\Xi^{0}_5$ \cite{NA49}. 

The HERA H1 collaboration reported the observation of
a narrow resonance in $D^{*-}p$ and $D^{*+} \bar p$ invariant mass combinations 
in inelastic electron-proton collisions~\cite{H1_Thetac}.
The resonance has a mass of 
3099$\pm$5 MeV and a measured 
Gaussian width of 12$\pm$ 3 MeV, compatible with 
the experimental resolution. The resonance was interpreted 
as a possible anti-charmed pentaquark baryon with a minimal constituent quark 
composition of $uudd\bar c$, together with the charge conjugate. 

Still, there are important difficulties  which need to be resolved
before any claims can be made about the existence of pentaquarks.
Perhaps most importantly, several high-energy accelerator 
laboratories have been unable to observe any corroborating 
signal, even in cases with high statistical precision \cite{high-energy}.  

In the following section we review the current experimental 
situation, and planned second generation experiments, with emphasis 
on the Jlab program. 

\section{CURRENT EXPERIMENTAL SITUATION}

The properties of the observed candidate pentaquark
signals involving a variety of probes and targets 
are presented in Table \ref{table:results}. 
\begin{table}[htb]
\caption{The $\Theta^+$ positive observations}
\label{table:results}
\begin{tabular}{|l|l|c|c|c|}
\hline
Exp.          &  Mass  & Width & Stat.     & Ref.        \\
              &  MeV   & MeV   & sign.     &            \\
\hline
LEPS          & 1540$\pm 10$ & $<$25  & 4.6$\sigma$        & \cite{LEPS_old}\\
DIANA         & 1539$\pm 2$ & $<$ 9   & 4.4$\sigma$         & \cite{DIANA}\\
CLAS(d)       & 1542$\pm 4$ & $<$21   & 5.2$\sigma$         & \cite{CLAS2}\\
CLAS(p)       & 1555$\pm 10$& $<$26   & 7.8$\sigma$         & \cite{CLAS3}\\
SAPHIR        & 1540$\pm 2$ & $<$25   & 4.8$\sigma$         & \cite{SAPHIR}\\
HERMES        & 1528$\pm 3$ & $17\pm 9$& 4--6$\sigma$         & \cite{HERMES}\\
ITEP          & 1533$\pm 5$ & $<$20   & 6.7$\sigma$         & \cite{ITEP}\\
SVD-2         & 1526$\pm 3$ & $<$24   & 5.6$\sigma$         & \cite{SVD}\\
COSY          & 1530$\pm 5$ & $<$18   & 4--6$\sigma$         & \cite{COSY}\\
ZEUS          & 1522$\pm 3$ & $8\pm 4$   & 4.6$\sigma$         & \cite{ZEUS}\\
\hline
\end{tabular}
\end{table}
The initial reported observation was at LEPS~\cite{LEPS_old}
in the reaction 
$\gamma+n\to K^+K^-n$ 
utilizing a $CH_2$ target.
The statistical significance of the signal 
was estimated as 4.6$\sigma$. 
\begin{figure}[htb]
\vspace{7pt}
\centerline{\includegraphics[width=10cm]{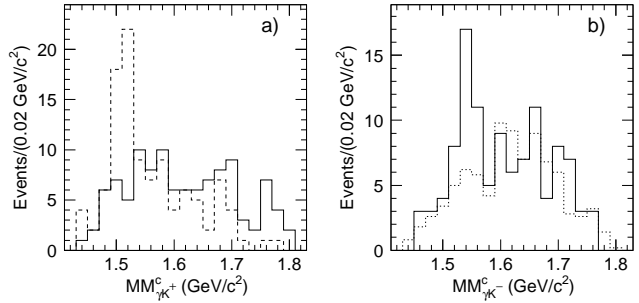}}
\caption{The first LEPS results. The solid line is the signal sample
and the dotted line is the background sample.
Left panel: the missing mass distribution $MM_{\gamma K^+}$
(the $\Lambda(1520)$ peak is seen).
Right panel: the missing mass distribution $MM_{\gamma K^-}$
(the $\Theta^+$ peak is seen).
Fermi motion corrections were 
applied to the data  to improve  the missing mass resolution.}
\label{fig:LEPS_old}
\end{figure}
The DIANA~\cite{DIANA} collaboration then reported  a possible
$\Theta^+$  signal in the charge-exchange reaction
$K^+Xe\to K^0pXe'$. 
This very narrow peak was observed 
 in the $K^0p$ effective mass spectrum
with 4.4 $\sigma$ statistical significance.
Soon thereafter 
the Jefferson Lab (JLab) CLAS collaboration  \cite{cite:CLAS} reported a positive 
result~\cite{CLAS1,CLAS2}  on a deuterium target. 

\begin{figure}[htb]
\vspace{7pt}
\includegraphics[width=10cm]{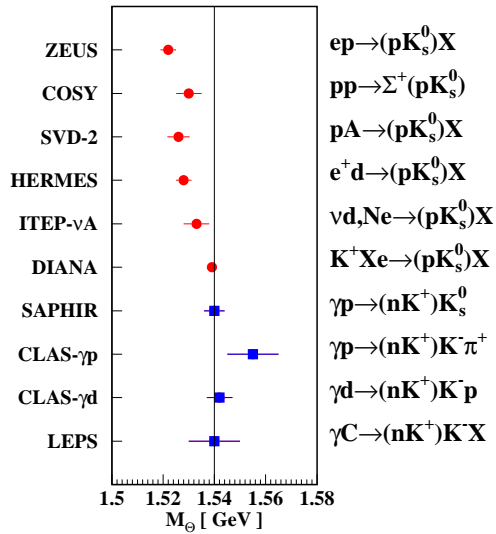}
\caption{The $\Theta^+$ mass measurements from various experiments. Note
the systematic difference in the measurements with
$nK^+$ and $pK^0$ in the final states.}
\label{fig:theta_mass}
\end{figure}

All of the above sightings were  
on nuclear targets. The first
positive observations reported on a proton target were in photoproduction
from  the JLab CLAS \cite{CLAS1,CLAS3} and the SAPHIR \cite{SAPHIR} 
collaborations. Since then there have been observations
with quasi-real photoproduction on deuterium \cite{HERMES}.
Positive observations were also made with neutrino interactions \cite{ITEP}, 
and hadron-hadron collisions \cite{SVD,COSY}
on various targets. 
Fig. \ref{fig:theta_mass}  summarizes
the published mass values of various experiments.

There are several noteworthy difficulties which have led
to reservations about accepting the existence of pentaquarks as being
established.
The statistical 
significance in every measurement is relatively small. The positions of the
observed peaks vary somewhat from case to case, with the masses of the 
$K^+n$ all higher than that for $K^0p$ \cite{cite:masses}
(see Fig. \ref{fig:theta_mass}).

A major problem, frequently mentioned,  is that the
high statistics searches at a number of high energy facilities
have not seen evidence of this state \cite{high-energy}. Clearly the observation of
pentaquark states in all high energy experiments would have been powerful
corroborations. However the very different kinematical and experimental conditions
between these high energy  semi-inclusive experiments
and the low energy 
exclusive  experiments 
do not allow a direct comparison. These null results
should not be taken as 
negation of the positive results.

Since the first published experimental observation there have been more
than 220 theoretical articles concerned with the properties of
pentaquarks (see \cite{Diakonov97,cite:theory} and references within).
At this time experimental knowledge of these properties 
is limited to a very modest determination of the mass range, $M=1522-1555$~MeV
and an upper limit on the width, $\Gamma < 10$~MeV.

Second generation
experimental programs at LEPS, JLab, COSY and KEK have been undertaken with
three main goals.
1. To corroborate earlier observations of the existence of pentaquark
states. 2. If they do exist, to determine their intrinsic properties, such as
mass, width, spin and parity.   3. To determine the reaction mechanism
for their production. 

\begin{figure}[htb]
\vspace{7pt}
\includegraphics[width=7cm, height=5cm]{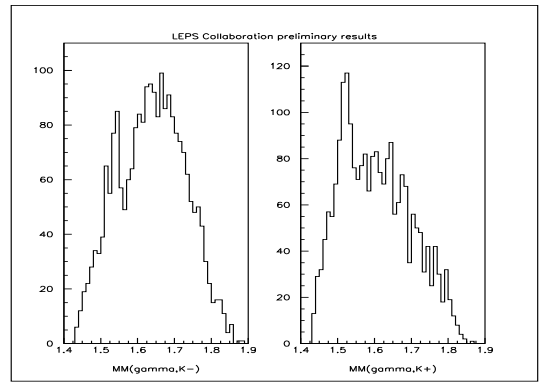}
\caption{
New LEPS preliminary data. The $K^-$ (left) and $K^+$ (right)
missing mass spectra for the
$\gamma+d\to K^+K^-X$ reaction are shown.
The $\Theta^+$ peak is seen at about
1.53 GeV on the left and the  $\Lambda(1520)$ peak is seen on the right.}
\label{fig:LEPS_new}
\end{figure}

The LEPS collaboration has already reported preliminary result \cite{LEPS_new} 
for the reaction
$\gamma n \to K^-K^+n$ using new data on a deuteron
target. The spectra   shown in Fig.~\ref{fig:LEPS_new}
appear to confirm their  initial observation.
This is considered one of the most compelling
positive results since it involves minimal kinematic cuts.
Followup experiments in COSY
to increase the statistics of their $pp\to \Sigma^+K^0_sp$ are scheduled for
2005. An extensive  experimental program has begun at JLab.
In the following we focus on the JLab initial published
results followed by a brief description of the ongoing and planned 
next generation experiments.   


\section{THE JEFFERSON LAB PROGRAM}

\subsection{Published Jefferson Lab Results}

The initial JLab results for deuteron and proton targets were obtained in the 
exclusive analysis of CLAS data that were recorded in experiments performed 
several  years earlier.

\begin{figure}[htb]
\vspace{7pt}
\includegraphics[width=7cm]{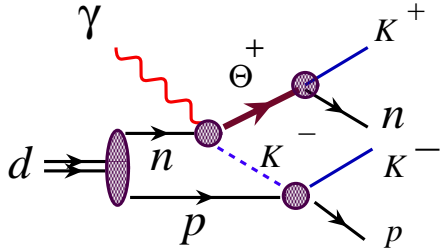}
\caption{A possible diagram for the $\Theta^+$ production from 
the deuterium target.}
\label{fig:FG1}
\end{figure}

\begin{figure}[htb]
\vspace{9pt}
\includegraphics[width=7cm]{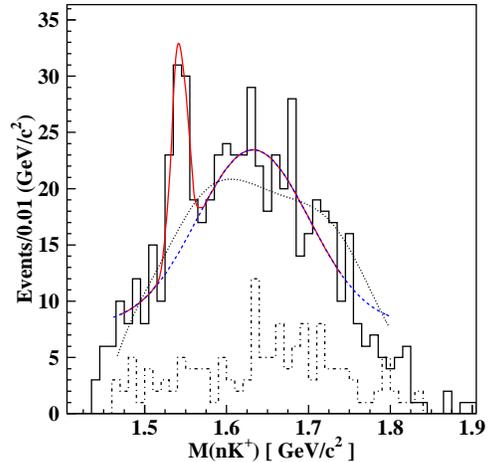}
\caption{The $M_{nK^+}$ invariant mass spectrum in the 
$\gamma d\to K^-K^+pn$ reaction obtained at CLAS.}
\label{fig:CLAS1}
\end{figure}

\vspace{0.1in}
\noindent{\bf Photoproduction on a Deuteron Target}.
The experiment was carried out with tagged photons having
maximum energy 3.1~GeV, incident on a 10 cm long deuterium target
\cite{CLAS2}.
The studied reaction  
$\gamma d\to K^-K^+pn$ was isolated by detecting the three charged particles
in the final state and selecting the neutron with the missing mass
technique.   
The $\Theta^+$  was reconstructed from the  $K^+n$ invariant mass.
Since detection of the proton was required to insure exclusivity,
a final state interaction involving the spectator proton was required in 
order to provide it with enough momentum to be detected.
The minimum proton momentum which was detected was 200 MeV/c.
A possible reaction diagram is shown in Fig.~\ref{fig:FG1}.
Cuts were applied to eliminate background  from known resonances:
the $\phi$(1020) mesons ( in the $K^+K^-$ decay mode) and
$\Lambda(1520)$ baryons (in the $pK^-$ decay mode).
The result is shown in Fig.~\ref{fig:CLAS1}. 
A peak is seen at a mass
$1542 \pm 5$ MeV.   The background under the
peak was estimated by two methods, a  Gaussian fit and Monte-Carlo 
simulation. The quoted statistical accuracy of the fitted peak was 
5.2 $\sigma$. An analysis of this data using a different technique finds
that the significance may not be as large as presented in the published
work. We expect a definitive answer from a much larger statistics data 
set that is currently being analyzed (see below).

\begin{figure}[htb]
\vspace{7pt}
\includegraphics[width=8cm]{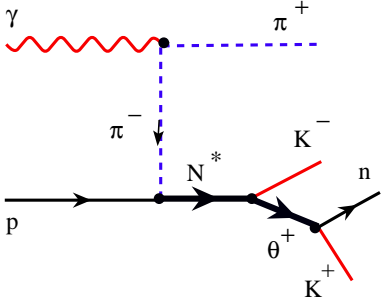}
\caption{A possible mechanism for the $\Theta^+$ production from a proton target.}
\label{fig:FG2}
\end{figure}

\begin{figure}[htb]
\vspace{7pt}
\includegraphics[width=7cm]{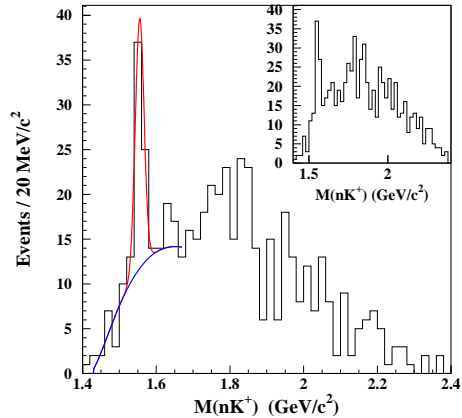}
\caption{The $M_{nK^+}$ invariant mass spectrum from CLAS in the 
$\gamma p\to \pi^+K^-K^+n$ reaction
 with the cut
$\cos\theta^*_{\pi^+}>0.8$ and $\cos\theta^*_{K^+}<0.6$, where
$\theta^*_{\pi^+}$ and $\theta^*_{K^+}$ are the
angles between the $\pi^+$ and $K^+$ mesons
and photon beam in the center-of-mass system.
The background function we used in the fit was obtained from the simulation.
The inset shows the $nK^+$ invariant mass spectrum  with only the 
$\cos\theta^*_{\pi^+}>0.8$ cut.
}
\label{fig:CLAS2_1}
\end{figure}

\vspace{0.1in}
\noindent{\bf Photoproduction on a Proton Target}.
The proton experiment used tagged photons with a maximum energy of
5.45 GeV, incident on a hydrogen target
\cite{CLAS3}.
The reaction studied
was  
$\gamma p\to \pi^+K^-K^+n$, 
with the neutron again identified
by missing mass. Several possible reaction mechanisms were considered,
$\gamma p\to K_0^* \Theta^+$ with a forward cut in the $K_0^*$  direction,
and $\gamma p\to \pi^+K^-\Theta^+$ with a forward cut in the $\pi^+$  direction,
illustrated in Fig.~\ref{fig:FG2}.
It was found that the signal was most evident in the 
channel with a forward going $\pi^+$.
The resulting mass spectrum is shown in Fig.~\ref{fig:CLAS2_1}.
A full partial wave analysis was performed on the reaction
$\gamma p\to \pi^+K^-K^+n$
to rule
out the possibility of a meson reflection in the $n K^+$ mass system
and to determine the 
background shape under the observed resonance structure.
The mass of the observed peak is $1555 \pm 10$ MeV. 
The statistical accuracy is quoted at about $7.8\pm 1$ $\sigma$.

\begin{figure}[htb]
\vspace{7pt}
\includegraphics[width=7cm]{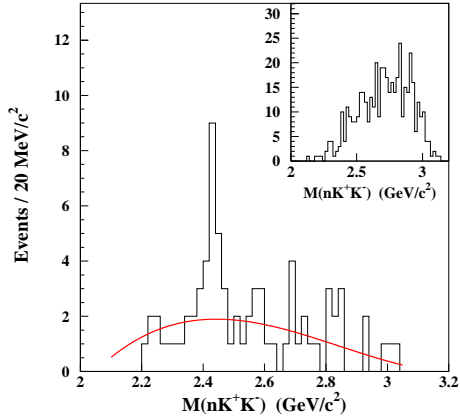}
\caption{The $M_{K^-nK^+}$ mass distribution for events selected from
the $\Theta^+$ peak.
The inset shows the distribution for the events outside the $\Theta^+$
region.
}
\label{fig:CLAS2_2}
\end{figure}

The invariant mass of the $\Theta^+K^-$ system was calculated,
yielding  the  distribution shown in Fig.~\ref{fig:CLAS2_2}. 
The observed
peak at a mass near 2.4 GeV is suggestive of a possible narrow resonance
state which decays into  $\Theta^+K^-$, as illustrated
in Fig.~\ref{fig:FG2}. 
Of course at this stage this is mere speculation.

\subsection{New Dedicated Experiments}
Based on the observations described here, the JLab CLAS and Hall A 
collaborations  have undertaken a series of experiments to try to corroborate 
the existence of pentaquarks. These are described in the following. 

\begin{figure}[htb]
\vspace{7pt}
\includegraphics[width=7cm]{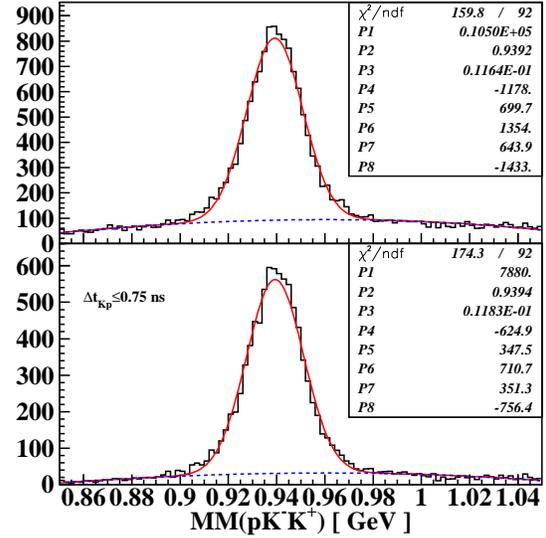}
\caption{
The missing mass spectra in the 
$\gamma d\to K^-K^+pX$ reaction.
The bottom panel shows the distribution with more stringent
vertex time cut. 
The neutron peak is clearly seen. (Preliminary CLAS data).
}
\label{fig:g10_n}
\end{figure}
\begin{figure}[htb]
\vspace{7pt}
\includegraphics[width=7cm]{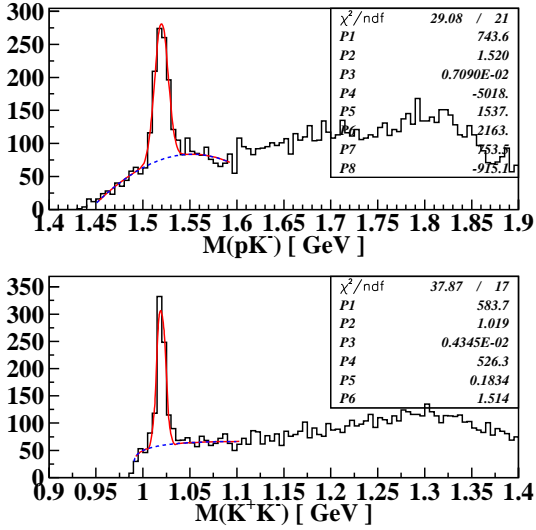}
\caption{
Top panel: the $M_{pK^-}$ invariant mass spectrum in the 
$\gamma d\to K^-K^+pn$ reaction, clearly showing the $\Lambda(1520)$
peak.
Bottom panel: the $M_{K^+K^-}$ invariant mass spectrum in the 
$\gamma d\to K^-K^+pn$ reaction exhibiting the $\phi$
peak. (Preliminary CLAS data).
}
\label{fig:g10_lambda}
\end{figure}

\vspace{0.05in}
\noindent{\bf Photoproduction of $\Theta^+$ from neutrons} ({\it g10})
\cite{cite:g10}.
This experiment, which was run during spring 2004, 
measured the exclusive reaction $\gamma d\to K^+K^-pn$ in order to
verify our  published result with an order of magnitude increase
in statistics. 
Data taking was successfully completed in May 2004. The experiment
utilized a maximum tagged photon beam energy of   3.6 GeV
with  two different magnetic field settings
for the CLAS spectrometer.
The target was shifted upstream
 25 cm from the nominal position to increase 
the acceptance in the forward direction. The run with lower 
magnetic field has 
increased acceptance  for forward going negative particles, which
allows us to perform an  analysis similar to  Spring-8 for inclusive 
reactions. The run
with  high magnetic field had the same geometrical 
acceptance and single track  resolution as the published CLAS result. 
The higher integrated luminosity was achieved 
by means of a longer target, improved trigger scheme, and 
longer data taking time. 
The total luminosity was about 50 $pb^{-1}$ compared with
about 2.5 $pb^{-1}$ for the originally published data.
The first pass of the detector calibration and performance check has 
already been completed. Preliminary examples of  missing mass and 
invariant mass 
distributions based on a small fraction of the statistics, shown in 
Fig.~\ref{fig:g10_n} and \ref{fig:g10_lambda}, give an indication of the
quality of the obtained data.
The missing mass spectra of the reaction  $\gamma d\to K^+K^-pX$ 
are shown in Fig.~\ref{fig:g10_n} for two different particle ID cuts.
The mass resolution ($\sigma=12~MeV/c^2$) and background under 
the neutron peak are in agreement with the previous measurements.
The $pK^-$ and $K^+K^-$ invariant mass spectrum in 
Fig. \ref{fig:g10_lambda} clearly shows the  $\Lambda(1520)$ baryon and $\phi$ 
meson, which are backgrounds that  have to  be removed in the final analysis.

\begin{table}[htb]
\caption{g10 experiment }
\label{table:g10_2}
\begin{tabular}{l|l|l}
\hline
Reaction       &  The $\Theta^+$  &  Detected   \\
               & decay mode       &  final state   \\
\hline
$\gamma d\to pK^+K^-n $       &  $nK^+$        & $pK^+K^-$       \\
$\gamma d\to \Lambda K^+n $   &  $nK^+$        & $p\pi^-K^+$     \\
$\gamma d\to \Lambda K^0 p$ &  $pK^0_S $     & $pp\pi^-$     \\
$\gamma d\to ppK^0_SK^-$      &  $pK^0_S $     & $p\pi^+\pi^-K^-$ \\
$\gamma ``n''\to K^+K^-n $    &  $nK^+$        & $K^+K^-$     \\
\hline
\end{tabular}\\[2pt]
\end{table}

Carefully controlled parallel independent analysis are being 
performed to insure maximum reliability of the results. 
The possible reaction and decay modes which  are the object
of the analysis   are shown in Table~\ref{table:g10_2}. 
Both decay modes, $pK^0$ and $nK^+$, are being analyzed in missing mass  and 
invariant mass distributions.
The results are expected to be released before the end of 2004.

\vspace{0.05in}

\noindent {\bf Search for the Ground and Excited States from  Protons} ({\it g11, super-g}) \cite{cite:g11,cite:superg}.
This consists of two experiments, each with an order of magnitude
more statistics than was available for our previously published result
on the proton~\cite{CLAS3} and preliminary (unpublished) 
lower energy data \cite{e1c}.

The first ({\it g11}, 
which utilized a tagged photon maximum energy of 
4 GeV successfully  completed data taking at the end of 
July 2004. The goal was primarily to check for the existence of
the $\Theta^+$ and other
possible members of the anti-decuplet on a proton target. 

The second ({\it super-g}) 
will be a comprehensive study of exotic baryons
from a proton target with a maximum photon energy of about 5.5 GeV,
with $5\times 10^7~s^{-1}$ tagged photons rate, and broad kinematic 
coverage for a variety of channels.
Its  goal is 
to corroborate the previously published results at a similar 
energy, to get information about the pentaquark spin by measuring  
decay angular distributions, and its reaction mechanism by measuring its
t-dependence. We will also investigate the possibility of an intermediate 
resonance near W=2.4 GeV as we reported in Ref.~\cite{CLAS3}. Another goal is to 
try to verify the existence of exotic cascades
which were reported by NA49~\cite{NA49}.
 The experiment is scheduled to run mid to late 2005.

To achieve the goals of these experiments it was necessary to design, 
fabricate and assemble a new longer target (40 cm) and new
detectors around the target to provide improved event triggering
and particle identification.

As mentioned, the first run ({\it g11}) was successfully completed.
There are $6.9\times 10^9$ events collected with
80~pb$^{-1}$ integrated luminosity.
The calibration of the different CLAS 
detectors, and checking of the CLAS performance is under way. We expect to have
the first physics result near the beginning of 2005. 
The reactions under study are: 
$\gamma p\to \bar K^0K^+n$, 
$\gamma p\to K^0\bar K^0p$, 
$\gamma p\to K^-\pi^+K^+n$, 
$\gamma p\to K^-\pi^+K^0p$, 
$\gamma p\to K^+\pi^+\Sigma^-$, 
$\gamma p\to K^+\pi^-\Sigma^+$,  and
$\gamma p\to K^-K^+p$.

\begin{figure}[htb]
\vspace{7pt}
\includegraphics[width=7cm]{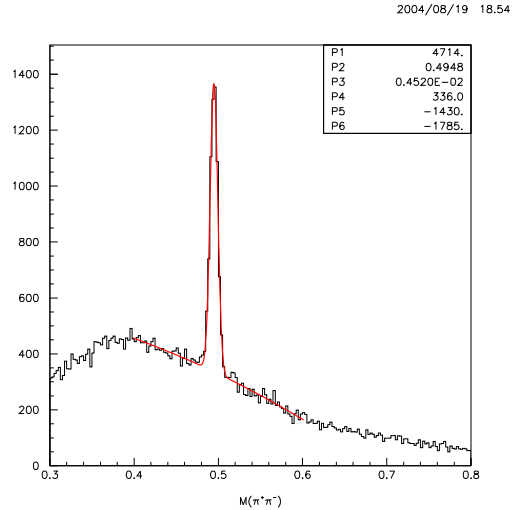}
\caption{
The $M_{\pi^+\pi^-}$ invariant mass spectrum in the 
$\gamma p\to \pi^+\pi^-K^+n$ reaction, showing the $K^0$
peak. (Preliminary CLAS data.)
}
\label{fig:g11_2}
\end{figure}
\begin{figure}[htb]
\vspace{7pt}
\includegraphics[width=7cm]{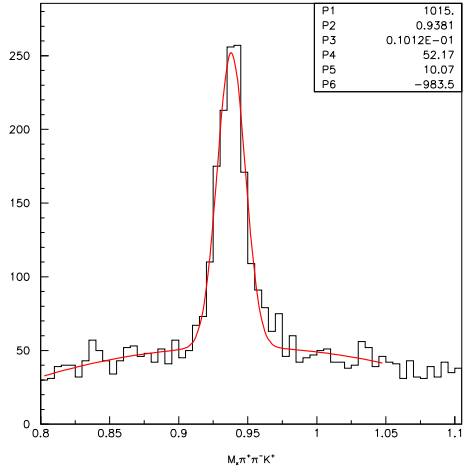}
\caption{
The missing mass spectrum in the 
$\gamma p\to K^0_SK^+X$ reaction.
The neutron peak is clearly seen. (Preliminary CLAS data).
}
\label{fig:g11_1}
\end{figure}
The performance of the CLAS setup is demonstrated in Figs.\ref{fig:g11_2} and
\ref{fig:g11_1}.
The missing mass resolution ($\sigma=10$ MeV/c$^2$) and invariant
mass resolution for the  $K^0$ meson ($\sigma=4.5$ MeV/c$^2$) is in 
agreement with spectrometer specifications. The background under 
the missing mass peaks will be reduced after the detector 
will be fully calibrated.

\vspace{0.05in}
\noindent{\bf The search for exotic cascades 
using an untagged virtual photon beam} ({\it eg3}) \cite{cite:eg3}.
The NA49 evidence for the  $\Xi^{--}_5$ 
and the $\Xi^0_5$ at a mass of
1.86 GeV \cite{NA49} was obtained from reconstruction of their decay products using 
the decays
$\Xi^{--}_5 \rightarrow \pi^- \Xi^{-}$ and $\Xi^0_5 \rightarrow \pi^+ \Xi^{-}$.
To date no other experiments have been able to confirm this observation.
{\it eg3} is
a new CLAS  experiment  searching for exotic cascades
using an untagged virtual photon beam.
The experiment is most sensitive to the  $\Xi^{--}_5$ and $\Xi^{-}_5$, by 
means of reconstruction
from their decay products. The main goal of the experiment will 
be to search
for $\Xi^{--}_5 \rightarrow \pi^- \Xi^-$, $\Xi^{-}_5 \rightarrow \pi^0 \Xi^-$ and
 $\Xi^{-}_5 \rightarrow \pi^- \Xi^0$. Other decay modes are detectable with 
lower sensitivity.

The experiment is designed to utilize the highest luminosity
possible with CLAS. It will 
use a 5.7 GeV electron beam incident on a deuterium target but without
detecting the scattered electron. The untagged photon
beam is necessary to achieve sufficient sensitivity to
the expected small cross sections.
The sequence
of weakly decaying daughter particles provides a powerful tool
to pick out the reactions of interest.
The sensitivity of the experiment is
expected to be 46 detected events per nb, and is scheduled to take data
in December 2004.

\vspace{0.05in}

\begin{figure}[htb]
\vspace{7pt}
\includegraphics[width=5cm,angle=-90]{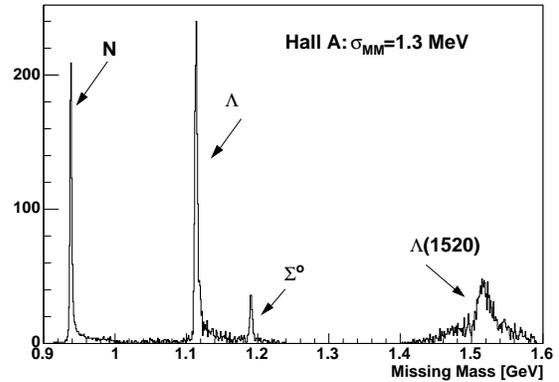}
\caption{Preliminary Hall A data.
The missing mass spectrum in the 
$e^-p\to e^-\pi^+X$ reaction. The neutron is seen (left peak).
The missing mass spectrum in the 
$e^-p\to e^-K^+X$ reaction (right peaks).
The $\Lambda$, $\Sigma^0$,and  $\Lambda(1520)$ peaks are seen.
}
\label{fig:bogdan}
\end{figure}

\noindent{\bf High Resolution search for $\Sigma^0_5$ and $\Theta^{++}$ in Hall A} \cite{cite:bogdan}.
The experiment has already completed data taking in June 2004.
It used a liquid hydrogen target and detected a forward $K^-$ (or $K^+$)
and scattered electron using the 2-arms of the high resolution
spectrometers of Hall-A \cite{HALL-A}. 
The reactions under study are $e^-p\to e^-K^+X$ and $e^-p\to e^-K^-X$.
The missing mass system
then had the characteristics of doubly-positive-charged ($K^-$ forward)
or neutral ($K^+$ forward) pentaquark states predicted in certain 
theoretical models. An excellent missing mass resolution provides the tool 
to search for  narrow-width states. The online spectrum is presented in 
Fig.~\ref{fig:bogdan}. Once again, data analysis is underway, with
results expected in Fall 2004.
  
\section{CONCLUSION}
Many laboratories are involved in the search for pentaquarks,
using a wide variety of beams, targets and final states. 
Several laboratories  have reported positive observation of the signal,
but with rather low statistical precision, while
others observe null results. Most, but not all, positive observations
were made at lower energy facilities and in exclusive reactions,
while almost all of the null results have been at high energy facilities
with inclusive reactions. Thus, at this time
the existence of a narrow pentaquark state is not fully confirmed.  

The question of whether pentaquarks exist can only be resolved
by a second generation of high statistics experiments. JLab is
in a unique position to address this. The experimental program which 
we have begun is expected to increase the integrated luminosity by 
more than an order of magnitude for each experiment
over what was  previously reported. Data for three of these experiments
is already in hand, and we expect  to present first results of 
the series of high-statistics experiments by the end of 2004.
Stay tuned!

\end{document}